\documentclass[prb,twocolumn,showpacs]{revtex4}
\usepackage{amsmath}
\usepackage{dcolumn}
\usepackage{graphicx}

\begin{document}

\title{A gapped $Z_{2}$ spin liquid phase with a $U(1)$ mean field ansatz: a
Bosonic resonating valence-bond description}
\author{Tao Li}
\affiliation{Department of Physics, Renmin
University of China, Beijing 100872, P.R.China}
\date{\today}

\begin{abstract}
Gapped $Z_{2}$ spin liquid as the simplest spin liquid has been proved to be the most difficult to realize in realistic models. Here we show that the frustration from a third-neighbor exchange $J_{3}$ on the spin-$\frac{1}{2}$ $J_{1}-J_{2}$ model on the square lattice may serve to stabilize such a long-sought state. We argue that a Bosonic RVB description is more appropriate than a Fermonic RVB description for such a gapped spin liquid phase. We show that while the mean field ansatz of the proposed state has a $U(1)$ gauge symmetry, the gauge fluctuation spectrum on it is actually gapped and the state should be understood rather as a $Z_{2}$ spin liquid state with topological order. The state is thus locally stable with respect to gauge fluctuation and can emerge continuously from a collinear Neel ordered phase.
\end{abstract}

\maketitle

Quantum spin liquids are exotic state of matter that support fractionalized excitations\cite{balents,palee,lhuillier}. The resonating valence bond(RVB) picture proposed more than four decades ago by Anderson remains the best way to envisage a quantum spin liquid\cite{RVB}. An RVB state is made of the coherent superposition of different singlet pairing patterns of the local spins on the lattice. The short-ranged RVB state is the first proposed and probably the simplest spin liquid state. It features a gapped spectrum in both spin triplet and spin singlet channel and possesses $Z_{2}$ topological order. On a two dimensional torus, it exhibits a typical four fold topological degeneracy. However, in spite of its seemingly innocent appearance, such a gapped $Z_{2}$ spin liquid state has been proved to be the most difficult to realize in realistic models.

Geometric frustration of interaction is the most important way to realize a quantum spin liquid. Numerical studies on frustrated quantum antiferromagnetic models have reported controversial evidences for the existence of gapped $Z_{2}$ spin liquids. For example, DMRG studies on both the spin-$\frac{1}{2}$ Kagome Heisenberg model and the $J_{1}-J_{2}$ model on the square lattice have reported gapped $Z_{2}$ spin liquid ground state in some parameter region\cite{white,jiang}. However, both claims suffer from uncertainty due to the smallness of the reported gap, especially that in the spin singlet channel, and are both challenged by further studies that report gapless ground state in the claimed parameter regions\cite{ran,liao,sheng}. This may suggest that these systems are still not sufficiently frustrated. 

In this paper, we will focus on the spin-$\frac{1}{2}$ $J_{1}-J_{2}$ model on the square lattice and ask if a gapped $Z_{2}$ spin liquid can be stabilized when we introduce additional frustration in the model. 
The $J_{1}{-}J_{2}$ model on the square lattice has been investigated by many people in the last
two decades\cite{doucot,gelfand,mila,schulz,vbs1,vbs2,capriotti,sushkov,mambrini,darradi,wangl,jiang,boson,sheng}.
At the classical level, the system is Neel ordered for
$J_{2}<0.5J_{1}$ with ordering wave vector $\mathrm{q}=(\pi,\pi)$. For $J_{2}>0.5J_{1}$, the Neel order is replaced by a stripy magnetic order with ordering wave vector $\mathrm{q}=(\pi,0)$ or $(0,\pi)$ after a first-order transition. Spin wave fluctuations will destroy the magnetic order in the intermediate region of $0.4J_{1} \lesssim J_{2} \lesssim 0.6J_{1}$. The nature of this disordered phase is under debate even after twenty years' intensive study. It is generally believed that some kind of spatial symmetry breaking will happen in this intermediate parameter region. 

More recently, a gapped $Z_{2}$ spin liquid phase has been reported by a DMRG study in the intermediate region of $0.41J_{1} \lesssim J_{2} \lesssim 0.62J_{1}$, which seems to emerge {\it continuously} from the Neel ordered phase for $J_{2}\lesssim 0.41J_{1}$\cite{jiang}. This claim, however, is challenged by a further DMRG study, which shows that the Neel ordered phase for $J_{2} \lesssim 0.44J_{1}$ is connected to a plaquatte valence bond solid state for $0.5J_{1}\lesssim J_{2}\lesssim 0.6J_{1}$, with a small gapless paramagnetic region in between\cite{sheng}. A continuous transition between the Neel ordered phase and a gapped symmetric spin liquid phase is also at odds with a well known effective field theory prediction\cite{read,read1,read2}, which claims that by quantum disordering a collinear antiferromagnetic order one inevitably encounter spatial symmetry breaking as a result of the instanton effect. 

The main purpose of this paper is to show that the above prediction of the effective field theory is not necessarily correct. Continuing a previous work on the $J_{1}-J_{2}$ model on the square lattice\cite{boson}, and equipped with a recent development on the $U(1)$ gauge field theory of a quantum antiferromagnet\cite{tri}, we show that by introducing the frustration from a third-neighbor exchange $J_{3}$, a fully gapped $Z_{2}$ spin liquid can develop continuously from a Neel ordered phase.  We show that although the spin liquid state we proposed has a mean field ansatz with a staggered $U(1)$ gauge symmetry, it should be understood as a $Z_{2}$ spin liquid state with intrinsic topological order and gapped gauge fluctuation spectrum. 
 
In this work we will describe the gapped spin liquid phase with the Bosonic RVB theory. As compared to the more commonly used Fermionic RVB theory, the Bosonic RVB theory has the following advantages for our purpose\cite{liang,sandvik,beach,xiu,bilayer}. Firstly, the Bosonic RVB theory can describe the gapped spin liquid phase and the magnetic ordered phase on the same footing and is thus particularly appropriate in the transition region between the two phases. The situation is totally different in the Fermionic RVB theory, in which one need to break the spin rotational symmetry by hand in the magnetic ordered phase. Secondly, the Bosonic RVB theory predicts that only the spin correlation beyond the correlation length would be significantly affected by the opening of the spinon gap. When the correlation length is much larger than the lattice scale, the local correlation would hardly be affected by the gap opening process, just as what we would expect for a spin liquid evolved continuously from the magnetic ordered phase\cite{liang,brvb}. The situation is again totally different in Fermionic RVB theory, in which gap opening in the spinon spectrum is necessarily accompanied by the appearance of symmetry breaking order parameters at the mean field level, which will affect the local correlation in the standard Landau fashion. This makes the Fermionic RVB state unlikely an appropriate description for a gapped spin liquid phase with incipient magnetic ordering instability.

In this paper, we consider the following model on the square lattice:
\begin{equation}
H=\sum_{i,j}
J_{i,j} \ \vec{\mathrm{S}}_{i}\cdot\vec{\mathrm{S}}_{j},\nonumber
\end{equation}
in which $J_{i,j}$ equals to $J_{1}$, $J_{2}$ and $J_{3}$ on first-, second- and third-neighbor bonds and is otherwise zero, $\vec{\mathrm{S}}_{i}$ denotes the spin operator on site $i$. 

In the Schwinger Boson representation\cite{schwinger}, the spin operator is written as $\vec{\mathrm{S}_{i}}=\frac{1}{2}\sum_{\alpha,\beta}
b^{\dagger}_{i,\alpha}\vec{\sigma}_{\alpha,\beta}b_{i,\beta}$,
in which $b_{i,\alpha}$ is a Boson operator that is subjected to the no double occupancy constraint of the form
$\sum_{\alpha}b^{\dagger}_{\alpha}b_{\alpha}=1$, $\vec{\sigma}$ is the
Pauli matrix. Such a representation has a built-in $U(1)$ gauge redundancy, as the spin operator is unaffected when we perform a $U(1)$ gauge transformation of the form $b_{i,\alpha}\rightarrow b_{i,\alpha}e^{i\phi_{i}}$, where $\phi_{i}$ is an arbitrary $U(1)$ phase. 

The Heisenberg exchange coupling can be written as
$\vec{\mathrm{S}}_{i}\cdot\vec{\mathrm{S}}_{j}=-\frac{1}{2}\hat{A}_{i,j}^{\dagger}\hat{A}_{i,j}=\frac{1}{2}\hat{B}_{i,j}^{\dagger}\hat{B}_{i,j}$,
where
$\hat{A}_{i,j}=b_{i\uparrow}b_{j\downarrow}-b_{i\downarrow}b_{j\uparrow}$ and
$\hat{B}_{i,j}=b_{i\uparrow}^{\dagger}b_{j\uparrow}+b_{i\downarrow}^{\dagger}b_{j\downarrow}$\cite{schwinger}. In the mean-field treatment, we replace $\hat{A}_{i,j}$ and $\hat{B}_{i,j}$
with their mean-field expectation value $A_{i,j}$ and $B_{i,j}$ and treat the no double occupancy constraint on average. We then have:
\begin{eqnarray}
   H_{\mathrm{MF}}=&-&\frac{1}{2}\sum_{i,j}\left(\Delta_{i,j}\hat{A}_{i,j}^{\dagger}+\mathrm{h.c.}\right)\nonumber\\
  &+&\frac{1}{2}\sum_{i,j}\left(F_{i,j}\hat{B}_{i,j}^{\dagger}+\mathrm{h.c.}\right)\nonumber\\
  &+&\lambda \sum_{i}
  \left(\sum_{\alpha} b_{i\alpha}^{\dagger}b_{i\alpha}-1\right),\nonumber
\end{eqnarray}
where $\Delta_{i,j}=J_{i,j}A_{i,j}$, $F_{i,j}=J_{i,j}B_{i,j}$, $\lambda$ is introduced to enforce the constraint on
average. The Bosonic RVB state can be constructed by Gutzwiller projection of the mean field ground state, which has the form of
\begin{equation}
|\mathrm{RVB}\rangle=\mathrm{P_{G}}[ \ \sum_{i,j}a(\mathrm{R}_{i}-\mathrm{R}_{j})b^{\dagger}_{i\uparrow}b^{\dagger}_{j,\downarrow}\ ]^{N/2}|0\rangle,\nonumber
\end{equation}
where $\mathrm{P}_{\mathrm{G}}$ is the Gutzwiller projector that enforces the
constraint of one Boson per site, $N$ is the number of lattice site.
The RVB amplitude $a(\mathrm{R}_{i}-\mathrm{R}_{j})$ is determined by the
parameters $\Delta_{i,j}$, $F_{i,j}$ and $\lambda$.

As a result of the $U(1)$ gauge redundancy of the Schwinger Boson representation, the
mean-field ansatz $H_{\mathrm{MF}}$ for a symmetric spin liquid should be invariant under the $U(1)$ gauge projective extension of the physical symmetry group\cite{wen,wangf,yang}.  The mean field ansatz that meets such a requirement can be classified by the projective symmetry group(PSG) technique, as is done for the square lattice models in [21] and [35].  As we argued in [21], when both $J_{2}$ and $J_{3}$ are subdominant as compared to $J_{1}$, the energetically most favorable Schwinger Boson mean field ansatz belongs to the so called zero-flux class and should satisfy the following rules. First, for
sites belonging to different sub-lattices, only a {\it real} $\Delta_{i,j}$
is allowed. Second, for sites in the same sub-lattice, only a {\it real}
$F_{i,j}$ is allowed. The structure of the mean field ansatz is the most transparent in the sublattice-uniform gauge, in which it takes the form of:
\begin{eqnarray}
F_{i,i+\vec{\delta}_{1}}&=&0, \ \ \ \  \ \Delta_{i,i+\vec{\delta}_{1}}=\Delta, \nonumber\\
F_{i,i+\vec{\delta}_{2}}&=&F, \ \ \ \ \ \Delta_{i,i+\vec{\delta}_{2}}=0, \nonumber\\
F_{i,i+\vec{\delta}_{3}}&=&F_{2x},\ \  \Delta_{i,i+\vec{\delta}_{3}}=0, \nonumber
\end{eqnarray}
where site $i$ belongs to the A sublattice, $\vec{\delta}_{\mu}$ (with $\mu=1,2,3$) denotes the
vectors connecting site $i$ to its neighbors, up to the third
distance. For sites in B
sublattice, the sign of $\Delta$ should be
reversed. An illustration of this mean field ansatz is given in Fig.1.  

The mean-field Hamiltonian can be solved most easily in the uniform gauge\cite{wangf,yang,gauge}, in which the mean field ansatz is manifestly translational invariant. The mean field Hamiltonian in this gauge is given by
\begin{eqnarray}
H_{\mathrm{MF}}=\sum_{\mathrm{k}
}\psi_{\mathrm{k}}^{\dagger} \left( {\begin{array}{*{20}c}
                                                                      \epsilon_{\mathrm{k}} &  \Delta_{\mathrm{k}} \\
                                                                     \Delta^{*}_{\mathrm{k}} & \epsilon_{\mathrm{k}}  \\
                                                                    \end{array} } \right)
                                                                    \psi_{\mathrm{k}},\nonumber
\end{eqnarray}
in which 
$\psi_{\mathrm{k}}^{\dagger}=(b_{\mathrm{k}\uparrow}^{\dagger},b_{\mathrm{-k}\downarrow})$,
$\epsilon_{\mathrm{k}}=\lambda+2F\delta(\mathrm{k})-2F_{2x}\eta(\mathrm{k})$, and
$\Delta_{\mathrm{k}}=2i\Delta\gamma(\mathrm{k})$. Here
$\delta(\mathrm{k})= \sin(\mathrm{k}_{x})\sin(\mathrm{k}_{y})$, $\eta(\mathrm{k})= (\cos(2\mathrm{k}_{x})+\cos(2\mathrm{k}_{y}))/2$,
$\gamma(\mathrm{k})=(\sin(\mathrm{k}_{x})+\sin(\mathrm{k}_{y}))/2$.
The mean-field spinon spectrum is given by
$E_{\mathrm{k}}=\sqrt{\epsilon_{\mathrm{k}}^{2}-\Delta_{\mathrm{k}}^{2}}$. 
The RVB amplitude derived from the mean-field ground state is
given by
\begin{equation}\label{eq:amplitude}
a(\mathrm{R}_{i}-\mathrm{R}_{j})=\frac{1}{N}\sum_{\mathrm{k}}\frac{\Delta_{\mathrm{k}}}{\epsilon_{\mathrm{k}}+E_{\mathrm{k}}}e^{i\mathrm{k}\cdot(\mathrm{R}_{i}-\mathrm{R}_{j})}\nonumber.
\end{equation}
It can be proved that the RVB amplitude
between sites in the same sub-lattice is identically zero.

\begin{figure}
\includegraphics[width=9cm]{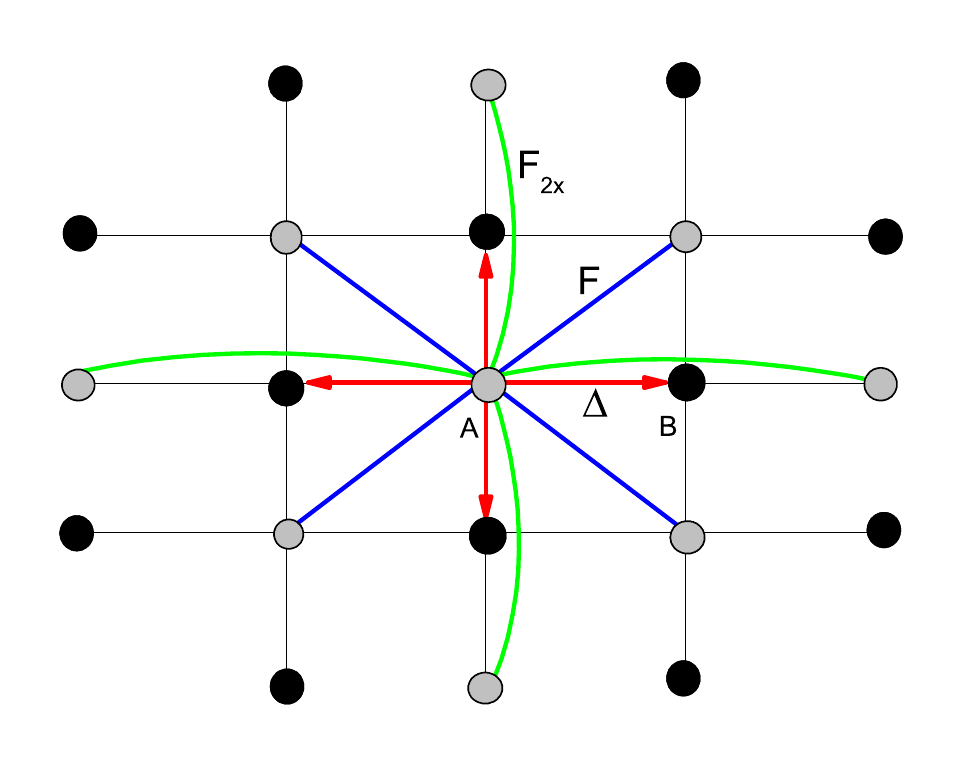}
\caption{\label{fig1}
(Color on-line) An illustration of the mean-field ansatz. Gray and dark dots denote sites in
sub-lattice A and B. }
\end{figure}

The mean-field parameters $\lambda$, $\Delta$, $F$ and $F_{2x}$ can be determined by solving the mean field self-consistent equations and the particle number equation. These equations have been solved earlier by Mila
and collaborators \cite{mila} when both $J_{3}$ and $F_{2x}$ are set to zero. They find that the spin liquid phase is preempted by the stripe magnetic ordered phase when $J_{2}\gtrsim0.6J_{1}$. We hope a gapped spin liquid phase can be stabilized by the additional frustration of $J_{3}$. 

The evolution of the spinon gap as determined by the solution of the self-consistent equations is shown in Fig.2. Here we restrict the parameters in the region $J_{2,3}/J_{1}<0.7$, since the mean field ansatz is constructed on the assumption that both $J_{2}$ and $J_{3}$ are sub-dominate. The spinon gap vanishes in the white region of the phase diagram, in which the Neel order with ordering wave vector $\mathrm{Q}=(\pi,\pi)$ emerges as a result of the condensation of the gapless spinon at momentum $\mathrm{Q}/2$. The Neel ordered phase is separated from the gapped spin liquid phase by a second order phase transition. The spinon gap in the spin liquid phase is found to increase linearly with the exchange coupling near the critical point(see Fig.3). The momentum where the minimal spinon gap is achieved is found to be always at or around $\mathrm{Q}/2$ in the parameter region considered. It can be seen that the critical value of $J_{2}$ to open the spinon gap decreases with the increase of $J_{3}$.  The reduction of the critical $J_{2}$ makes the gapped spin liquid phase to have better chance to survive the competition with the stripe ordered phase, since the frustration effect caused by $J_{3}$ is similar in both the Neel ordered and the stripe ordered phase. Here we do not attempt to make a thorough comparison of mean field energies for all possible symmetry breaking phases\cite{mean}, but choose to believe that a finite $J_{3}$ may stabilize a gapped spin liquid phase somewhere in the phase digram in proximity to the Neel ordered phase, before it is preempted by some unknown symmetry breaking phase.

\begin{figure}
\includegraphics[width=9cm]{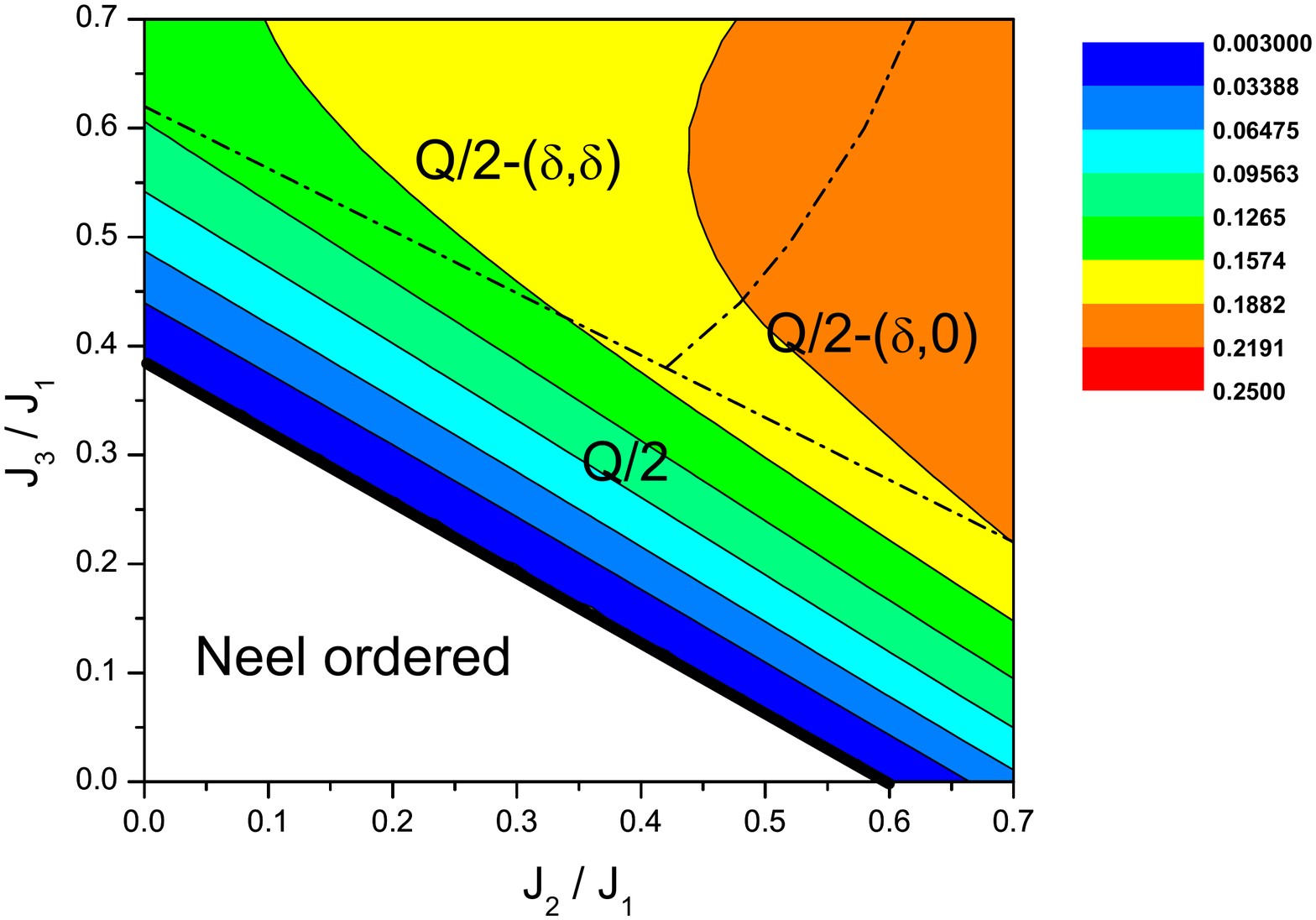}
\caption{\label{fig2}
(Color on-line) The mean field phase diagram of the $J_{1}-J_{2}-J_{3}$ model on the square lattice. The dark-thick line denotes a second order phase transition between the Neel ordered phase and the gapped spin liquid phase. The spinon gap is indicated by color scale. The dash-dotted lines denote the transition point where the spinon gap minimum change its momentum. Here $Q=(\pi,\pi)$ is the antiferromagnetic ordering wave vector, $\delta$ is a continuously varying momentum shift.}
\end{figure}

\begin{figure}
\includegraphics[width=9cm]{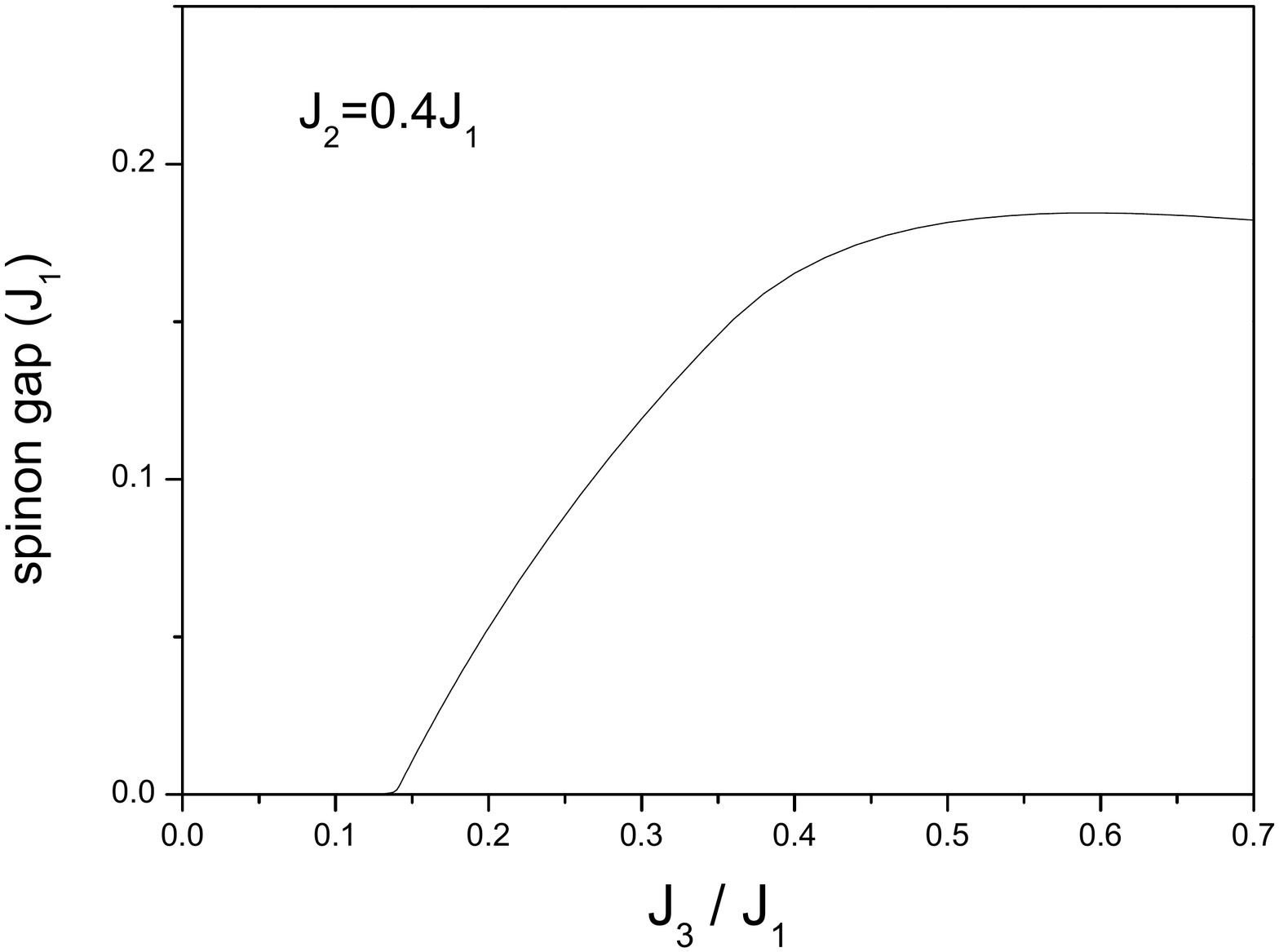}
\caption{\label{fig3}
The spinon gap as a function of $J_{3}$ for $J_{2}=0.4J_{1}$.}
\end{figure}

We now go beyond the saddle point approximation and construct a low energy effective theory for the fluctuation around the saddle point. The fluctuation in the magnitude of the RVB parameters, namely, $|\Delta|$ , $|F|$ and $|F_{2x}|$, are all gapped and can be neglected at low energy, since these fluctuations are not related to any symmetry. On the other hand, the saddle point action of the gapped spin liquid phase is manifestly invariant under a staggered $U(1)$ gauge transformation of the form: $b_{i,\alpha}\rightarrow b_{i,\alpha}e^{i\phi}$ for $i\in A$ and  $b_{j,\alpha}\rightarrow b_{j,\alpha}e^{-i\phi}$ for $j\in B$, since $\Delta_{i,j}$ is nonzero only for sites in different sub-lattices and $F_{i,j}$ is nonzero only for sites in the same sublattice\cite{read,read1,read2,wangf,yang}. This staggered $U(1)$ gauge symmetry is a reflection of the collinearity of the Neel ordered phase when the spinon condenses. At the Gaussian level, such a $U(1)$ gauge symmetry in the saddle point action would imply a gapless staggered $U(1)$ gauge field in the long wave length limit. According to a well known argument\cite{read,read1,read2}, when the spinon is fully gapped, the singular gauge field configuration called instanton in such a gapless staggered $U(1)$ gauge field will proliferate, which will result in spontaneous spatial symmetry breaking and the development of valence bond solid order.  A fully symmetric gapped $Z_{2}$ spin liquid phase thus can not emerge continuously from a collinear Neel ordered phase.

Here we show the above argument is incorrect and that a fully symmetric gapped $Z_{2}$ spin liquid phase can emerge continuously from a collinear Neel ordered phase. The key point of our reasoning is the observation that the fluctuation in the Lagrange multiplier, which enforces the no double occupancy constraint on the Schwinger Bosons, can not be treated at the perturbative level. In fact, one should integrate out the Lagrange multiplier exactly to find the correct low energy effective action for the emergent gauge fluctuation. A similar argument has been recently put forward by the present author in the study of the $U(1)$ spin liquid phase with a large spinon Fermi surface on the triangular lattice\cite{tri}. In that work, we find that the effective action of the gauge fluctuation takes a totally different form from that in the Gaussian effective theory when the Lagrange multiplier is exactly integrated out. Here we follow the same reasoning to derive the low energy effective action for the gauge fluctuation in the gapped Bosonic spin liquid.

We first rewrite the partition function of the system in the functional path integral representation in terms of the Schwinger Bosons. It takes the form of\cite{schwinger,read}
\begin{eqnarray}
Z=\int \prod_{i,\alpha,\tau} \mathcal{D} b_{i,\alpha}(\tau) \mathcal{D}b^{\dagger}_{i,\alpha}(\tau)\mathcal{D}\lambda_{i}(\tau)\ e^{-S},\nonumber
\end{eqnarray} 
in which 
\begin{eqnarray}
S&=&\int_{0}^{\beta} d\tau [\  \sum_{i}b^{\dagger}_{i,\alpha}(\tau)\partial_{\tau}b_{i,\alpha}(\tau)+H\nonumber\\
&+&\ \ \ i\sum_{i}\lambda_{i}(\tau)(b^{\dagger}_{i,\alpha}(\tau)b_{i,\alpha}(\tau)-1)\   ].\nonumber
\end{eqnarray}
Here $\lambda_{i}(\tau)$ is the Lagrange multiplier introduced to enforce the no double occupancy constraint on the Schwinger Bosons. $H=1/2\sum_{i,j}J_{i,j}\hat{B}^{\dagger}_{i,j}\hat{B}_{i,j}$ or $H=-1/2\sum_{i,j}J_{i,j}\hat{A}^{\dagger}_{i,j}\hat{A}_{i,j}$ is the Hamiltonian written in terms of the Schwinger Bosons, depending on the decoupling channel we will adopt.  Introducing the complex Hubbard-Stratonovich field $A_{i,j}$ or $B_{i,j}$ to decouple the Hamiltonian, the partition function can be written as
\begin{eqnarray}
Z=C \int \prod \mathcal{D} b \mathcal{D}b^{\dagger} \mathcal{D}\lambda  \mathcal{D}\varphi \mathcal{D}\bar{\varphi}  \ e^{-S'},\nonumber
\end{eqnarray} 
here we have omitted the arguments of the Boson fields and the auxiliary fields for clarity, $\varphi_{i,j}(\tau)$ is the Hubbard-Stratonovich field defined on the bond connecting site $i$ and $j$. Depending on the bond type, it equals to either $A_{i,j}(\tau)$ or $B_{i,j}(\tau)$. The action $S'$ is given by
\begin{eqnarray}
S'&=&\int d\tau \sum_{i}b^{\dagger}_{i}(\partial_{\tau}+i\lambda_{i})b_{i}\nonumber\\
&-&\frac{J_{1}}{2}\sum_{<i,j>_{1}}(\ \bar{A}_{i,j}\hat{A}_{i,j}+A_{i,j}\hat{A}^{\dagger}_{i,j}\ )\nonumber\\
&-&\frac{iJ_{2}}{2}\sum_{<i,j>_{2}}(\ \bar{B}_{i,j}\hat{B}_{i,j}+B_{i,j}\hat{B}^{\dagger}_{i,j}\ )\nonumber\\
&-&\frac{iJ_{3}}{2}\sum_{<i,j>_{3}}(\ \bar{B}_{i,j}\hat{B}_{i,j}+B_{i,j}\hat{B}^{\dagger}_{i,j}\ )\nonumber\\
&+&\frac{1}{2}\sum_{i,j}J_{i,j}\bar{\varphi}_{i,j}\varphi_{i,j}-i\sum_{i}\lambda_{i}\ \ ,\nonumber
\end{eqnarray} 
in which $<i,j>_{1}$, $<i,j>_{2}$ and $<i,j>_{3}$ denote first-, second- and third-neighbor bonds respectively.

As usual, we neglect the fluctuation in the magnitude of the auxiliary fields $A_{i,j}$ and $B_{i,j}$ at low energy, which are assumed to be both gapped. The phase of the auxiliary fields and the Lagrange multiplier are then the only degree of freedoms in the low energy effective theory, which are to be interpreted as the spatial and temporal component of a compact $U(1)$ gauge field. Such an emergent gauge field has no intrinsic dynamics of its own. An effective dynamics of the gauge fluctuation can be generated by integrating out the Schwinger Bosons. As we have shown in [25], it is crucial to integrate out the Lagrange multiplier exactly, which enforce the no double occupancy constraint on the slave particles, to derive the correct effective action for the spatial component of the emergent $U(1)$ gauge field. 

When the Lagrange multiplier is integrated out exactly, the coupling of the gauge field to all gauge non-invariant quantities should vanish identically, since the latter will inevitably violate the no double occupancy constraint. What survives the integration is then the coupling between the gauge invariant fluxes and the corresponding gauge invariant loop operators\cite{tri}. For example, if we only consider RVB parameters on first-neighbor bonds, then to the lowest order of $|\Delta|$, the coupling between the gauge field and the Schwinger Bosons is given by 
\begin{eqnarray}
H'_{1}=i\gamma_{1} \sum_{[i,j,k,l]} \Phi_{[i,j,k,l]} (\hat{A}_{i,j}\hat{A}^{\dagger}_{j,k}\hat{A}_{k,l}\hat{A}^{\dagger}_{l,i}-\mathrm{h.c.}),\nonumber 
\end{eqnarray}
in which the sum is over all elementary plaquettes of the square lattice. $[i,j,k,l]$ denotes a plaquette composed of the site $i$, $j$, $k$ and $l$(see Fig4(a) for an illustration). $\Phi_{[i,j,k,l]}$ is the gauge flux enclosed in this plaquette and is defined as $\Delta^{*}_{i,j}\Delta_{j,k}\Delta^{*}_{k.l}\Delta_{l.i}=|\Delta|^{4}e^{i\Phi_{[i,j,k,l]}}$. $\gamma_{1}$ is a coupling constant that is proportional to $|\Delta|^{4}$. A detailed derivation of a similar result for the $U(1)$ spin liquid on the triangular lattice can be found in our recent paper\cite{tri}. 

If we include the second- and the third-neighbor RVB parameters, we can define more gauge invariant fluxes and the corresponding gauge invariant loop operators. For example, we can generate the following coupling by combining first- and second-neighbor RVB parameters 
\begin{eqnarray}
H'_{2}=i\gamma_{2} \sum_{[p,q,r]} \Phi'_{[p,q,r]} (\hat{A}^{\dagger}_{p,q}\hat{A}_{q,r}\hat{B}_{r,p}-\mathrm{h.c.}),\nonumber 
\end{eqnarray}
in which $[p,q,r]$ is a triangular plaquette as illustrated in Fig.4(b). $\Phi'_{[p,q,r]}$ is the gauge flux enclosed in the plaquette $[p,q,r]$. It is defined as $\Delta_{p,q}\Delta^{*}_{q,r}F^{*}_{r,p}=|\Delta|^{2}|F|e^{i\Phi'_{[p,q,r]}}$. The coupling constant $\gamma_{2}$ is proportional to $|\Delta|^{2}|F|$. Obviously, there are infinite number of such gauge invariant fluxes and the corresponding loop operators. However, one can show that all gauge invariant fluxes can be generated by combination of five kinds of elementary gauge fluxes. In Fig.4(b)-4(f), we illustrate the five kinds of loops on which these elementary gauge fluxes are defined.

\begin{figure}
\includegraphics[width=12cm]{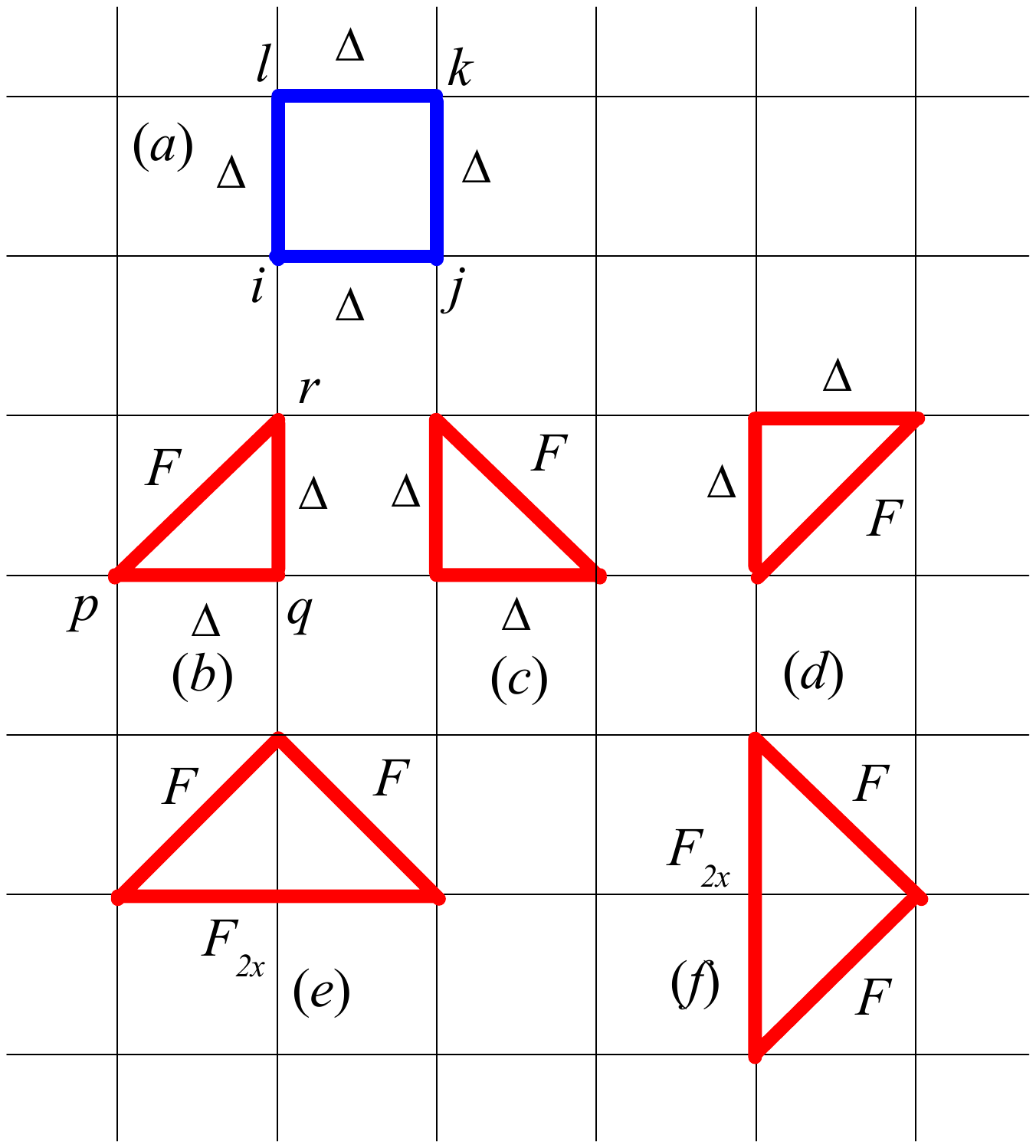}
\caption{(Color on-line) The various loop operators defined on the square lattice. (a) A loop operator defined on the plaquette $[i,j,k,l]$. (b)-(f) The loops used to defined the five kinds of elementary gauge fluxes.}
\end{figure}

The dynamics of the gauge fluxes can be determined by studying the spectrum of the loop operators, since the two are linearly coupled with each other. Since the Bosonic spinon is fully gapped in the spin liquid phase, we expect the fluctuation of the loop operators to be also gapped. This implies that the gauge fluctuation in the spin liquid phase is fully gapped, although the mean field ansatz of the spin liquid phase possesses the staggered $U(1)$ gauge symmetry. The spin liquid phase should thus be understood as a gapped $Z_{2}$ spin liquid phase with topological order. The gapped nature of the gauge fluctuation also implies that the spin liquid phase is locally stable with respect to gauge fluctuation. A continuous transition between the collinear Neel ordered phase and a gapped $Z_{2}$ spin liquid phase with the full symmetry is thus possible.

In conclusion, we find that the frustration from a third-neighbor exchange may stabilize a fully symmetric gapped $Z_{2}$ spin liquid phase in the  spin-$\frac{1}{2}$ $J_{1}-J_{2}$ model on the square lattice. We argue that such a gapped spin liquid phase is better described by the Bosonic, rather than the Fermionic RVB theory. We find although the mean field ansatz of such a spin liquid phase has a staggered $U(1)$ gauge symmetry, it should be understood as a system with gapped gauge fluctuation spectrum and $Z_{2}$ topological order. We argue that such a gapped spin liquid phase is locally stable against gauge fluctuation and can have a continuous transition with the collinear Neel ordered phase. This claim is in strong contradiction with previous field theory predictions based on Gaussian approximation on the Lagrange multiplier, in which the gapped spin liquid phase is argued to be unstable against spatial symmetry breaking as a result of the instanton effect. Our result illustrate again that it is crucial to enforce the no double occupancy constraint on the slave particles {\it exactly} to construct the correct low energy effective theory for the emergent gauge field in a quantum spin liquid. Arguments based solely on the gauge symmetry of the mean field ansatz can be misleading.

The author acknowledges the support from NSFC Grant No. 11674391, 973 Project No. 2016YFA0300504, and the research fund from Renmin University of China.


\begin{thebibliography}{99}
\bibitem{brvb}In both the Bosonic and the Fermionic RVB theory, the RVB amplitude is given by $a(\mathrm{R}_{i}-\mathrm{R}_{j})=\frac{1}{N}\sum_{\mathrm{k}}\frac{v_{\mathrm{k}}}{u_{\mathrm{k}}}e^{i\mathrm{k}\cdot(\mathrm{R}_{i}-\mathrm{R}_{j})}$. However, the Fourier component $\frac{v_{\mathrm{k}}}{u_{\mathrm{k}}}$ has distinct property in the two theories. While it diverges at the gapless point in the Fermionic RVB theory, it approaches unity in the Bosonic RVB theory. This difference can be traced back to the difference in the normalization condition of $u_{\mathrm{k}}$ and $v_{\mathrm{k}}$ in the two theories.
\bibitem{gauge} The gauge transformation
relating the uniform gauge and the sub-lattice uniform gauge is given by
$(-1)^{[\frac{i_{x}+ i_{y}}{2}]}$, in
which $[r]$ means the largest integer that is not greater than $r$.
\bibitem{mean}It can be very misleading to compare the mean field energies of different symmetry breaking phases, since the effect of Gutzwiller projection can be very different in these phases.
\end{thebibliography}
\end{document}